\documentstyle[prl,aps,multicol,epsf]{revtex}
\begin{document}
\draft
\title{Multiscaling in Models of Magnetohydrodynamic Turbulence}
\author{Abhik Basu$^1$, Anirban Sain$^1$, Sujan K. Dhar$^2$,
and Rahul Pandit$^1$ \cite{byjnc}}
\address{$^1$Department of Physics, and $^2$Supercomputer
Education and Research Center, Indian Institute of Science,\\
Bangalore - 560012, India}
\maketitle
\date{\today}
\begin{abstract}
From a direct numerical simulation of the MHD equations we show,
for the first time, that velocity and magnetic-field structure 
functions exhibit multiscaling, extended self similarity (ESS), and
generalized extended self similarity (GESS). We also propose a new
shell model for homogeneous and isotropic MHD turbulence, which
preserves all the invariants of ideal MHD, reduces to a
well-known shell model for fluid turbulence for zero magnetic
field, has no adjustable parameters apart from Reynolds numbers, and
exhibits the same multiscaling, ESS, and
GESS as the MHD equations. We also study dissipation-range asymptotics and the
inertial- to dissipation-range crossover. 
\end{abstract}


\begin{multicols}{2}
The extension of Kolmogorov's work (K41)
\cite{k41} on fluid turbulence to
magnetohydrodynamic (MHD) turbulence yields \cite{mont}
simple scaling for velocity ${\bf{v}}\/$ and magnetic-field
${\bf{b}}\/$ structure functions, 
for distances $r\/$ in the {\em inertial range} 
between the forcing scale $L\/$ and the dissipation scale $\eta_d\/$. 
Many studies have shown that there are multiscaling
corrections to K41 scaling in fluid turbulence
\cite{ansr}. 
 Solar-wind data
\cite{krug} and recent shell-model studies
\cite{biskamp,carbone} for MHD turbulence yield similar multiscaling. 
We
elucidate this for homogeneous, isotropic MHD turbulence, in the
absence of a mean magnetic field, by presenting the first
evidence for such multiscaling in a pseudospectral study of
the MHD equations in three dimensions (henceforth
$3d$MHD). We also propose a new shell model with no adjustable parameters
(apart from Reynolds numbers)
which displays this multiscaling
and reduces
to the 
Gledzer-Ohkitani-Yamada (GOY) shell model \cite{goy,kada} for $3d\/$
fluid turbulence when $\bf b=0$. To extract multiscaling exponents 
we develop the
ideas of extended self similarity
(ESS) \cite{benzi1,Dhar} and generalised extended self
similarity (GESS) \cite{Dhar,benzi2} in both real and
wave-vector (henceforth $k$) spaces, that have been used in
fluid turbulence \cite{benzi1}-\cite{benzi2}.

We use the structure
functions ${\cal{S}}_p^a = \langle |{\boldmath{a}}({\bf{x}}+{\bf{r}})
- {\boldmath{a}}({\bf{x}})| ^p \rangle$, where ${\boldmath{a}}\/$ can be 
${\bf{v}}$, ${\bf{b}}$, or one of the Els\"{a}sser variables
${\bf{Z}}^{\pm} = {\bf{v}} \pm {\bf{b}}\/$, ${\bf{x}}$ and
${\bf{r}}$ are spatial coordinates, and the angular brackets
denote an average in the statistical steady state.
${\cal S}_p^a \sim
r^{\zeta_p^a}\/$ at high fluid and magnetic Reynolds numbers
$Re\/$ and  $Re_b\/$, respectively, and for the inertial
range $20\eta_d \lesssim r << L\/$.
The extension \cite{mont} of K41
to {\em homogeneous, isotropic} MHD turbulence {\em
with no mean magnetic field} yields $\zeta_p^a = p/3\/$. 
Shell models \cite{biskamp,carbone} and solar-wind data
\cite{krug} have obtained multiscaling in MHD turbulence, 
i.e., $\zeta_p^a = 
p/3 - \delta \zeta_p^a\/$, with $\delta \zeta_p^a > 0\/$ and
$\zeta_p^a\/$ nonlinear, monotonically increasing functions of
$p\/$. Work on fluid turbulence suggests 
\cite{ansr} an
extension of the apparent inertial range if we use ESS
\cite{benzi1} and GESS \cite{benzi2}: Thus with ESS, in
which $\zeta_p^a/\zeta_3^a\/$ follows from ${\cal{S}}_p^a \sim
[{\cal{S}}_3^a]^{\zeta_p^a/\zeta_3^a}\/$, we should expect by
analogy that it extends down
to $r \simeq 5\eta_d$ (as exploited in some MHD
shell models \cite{biskamp,carbone}). In GESS,
which employs ${\cal{
G}}_p^a(r) \equiv {\cal{S}}_p^a (r)/[{\cal{S}}_3^a (r)]^{p/3}$
and postulates a form ${\cal{G}}_p^a (r) \sim [{\cal{G}}_q^a
(r)]^{\rho^a_{pq}}$, with $\rho^a_{pq}=[\zeta_p^a - p\zeta_3^a/3]/[\zeta_q^a 
-q\zeta_3^a/3]$, it has been suggested \cite{benzi2} for fluid
turbulence that the apparent inertial range is extended to the
lowest resolvable $r\/$; however,  $k$-space 
GESS  \cite{Dhar} shows a crossover from
inertial- to dissipation-range asymptotic behaviors. 
GESS has not been
used in MHD turbulence so far.

Our studies 
yield many interesting results: 
The multiscaling exponents we obtain from $3d$MHD and our shell model
studies agree (Figs. 1a and 1b) and 
$\zeta_p^b>\zeta_p^{Z^+}$ \raisebox{-3pt}
{$\stackrel{>}{\sim}$} $\zeta_p^{Z^-}>\zeta_p^v\/$.
$\zeta_p^b\/$ lie close to the
She-Leveque (SL) prediction \cite{shelev} for fluids
($\zeta_p^{SL} =
p/9 + 2[1 - (2/3)^{p/3}]\/$), but $\zeta_p^v\/$
lie below it (Fig. 1c) \cite{foot3}.  These differences between velocity
and magnetic-field exponents are also mirrored in differences in
the probability distribution functions (Fig. 1d) 
for $\delta v_{\alpha}({\bf{r}}) = {v_{\alpha}}({\bf{x}}+{\bf{r}}) - 
{v_{\alpha}}({\bf{x}})$ and $\delta b_{\alpha}({\bf{r}}) =
{b_{\alpha}}({\bf{x}}+{\bf{r}})
- {b_{\alpha}}({\bf{x}})$. ESS
works both with real- and $k$-space structure
functions (Fig. 2). To study the
latter we 
postulate $k$-space ESS (for real-space structure functions we
use ${\cal{S}}\/$ and ${\cal{G}}\/$ and for their
$k-$space analogs ({\em not} Fourier transforms) 
$S\/$ and $G\/$):
\begin{eqnarray}
S_p^a &\equiv& \langle |{\boldmath{a}}({\bf{k}})|^{p}
\rangle \approx A^a_{Ip} (S_3)^{\zeta^{'a}_p}, \;\;L^{-1} \ll k
\lesssim 1.5 k_d, \nonumber \\
S_p^a &\equiv& \langle |{\boldmath{a}}({\bf{k}})|^{p}
\rangle \approx A^a_{Dp} (S_3)^{\alpha_p^a}, \;\;1.5 k_d \lesssim k \ll
\Lambda, 
\end{eqnarray}
where $A^a_{Ip}\/$ and $A^a_{Dp}\/$ are, respectively, nonuniversal
amplitudes for inertial and dissipation ranges and
$\Lambda^{-1}\/$ the (molecular) length at which hydrodynamics
breaks down (cf. \cite{Dhar} for fluid turbulence).  We find that
$\alpha^a_p \neq \zeta^{'a}_p\/$. In
our shell model $\zeta^{'a}_p = \zeta_p^a\/$, but our data for
$3d$MHD suggest $\zeta^{'a}_p = 2(\zeta^a_p + 3
p/2)/11\/$ (i.e., $S_p^a(k) \sim k^{-(\zeta_p^a + 
3p/2)}\/$ in the inertial range \cite{foot}); the 
difference arises because of phase-space factors \cite{Dhar}.
\end{multicols}
\vspace*{-2.5cm}
\begin{figure}[htb]
\epsfysize=16cm
\centerline{\epsfbox{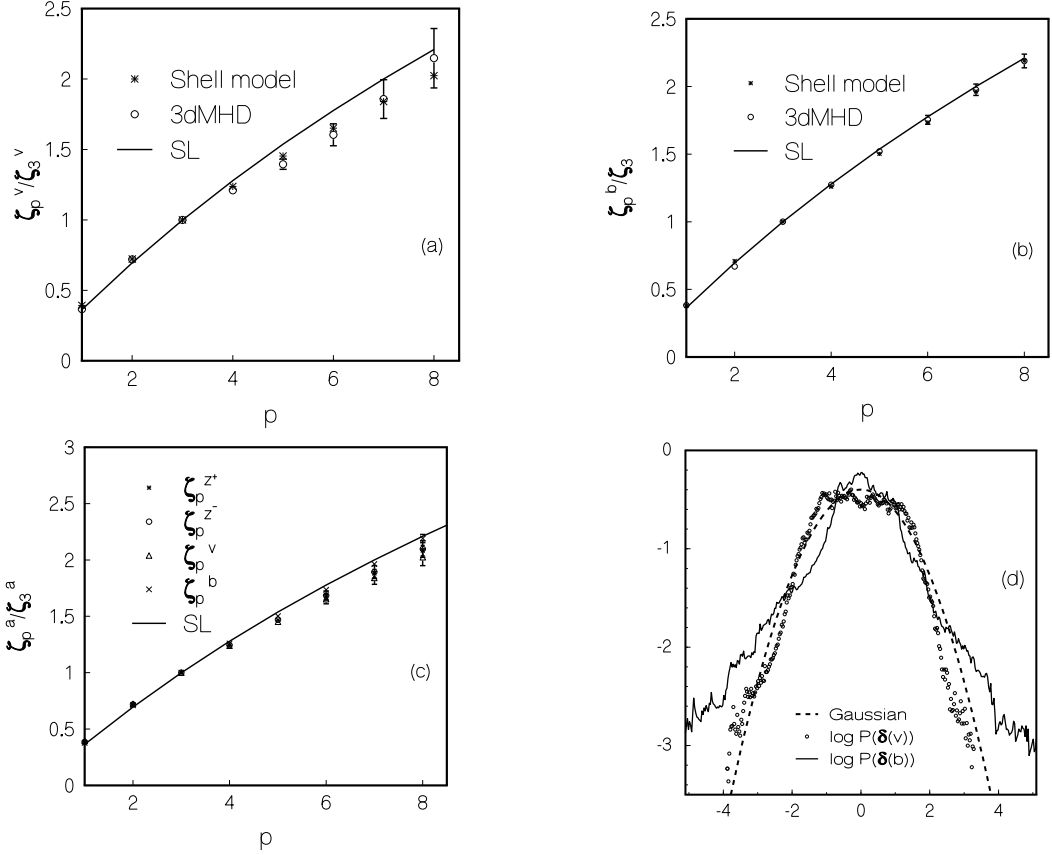}}
\vspace*{-3cm}
\caption{(a)-(c)Inertial-range exponents versus $p\/$ from typical $3d$MHD and
shell-model runs (Table 1) and their comparison with the SL formula:
 (a) $\zeta_p ^v/\zeta_3 ^v$,
(b) $\zeta_p ^b/\zeta_3 ^b$, and
(c) $\zeta_p ^v$, $\zeta_p 
^b$, $\zeta_p ^{z^+}$, and $\zeta_p ^{z^-}$ from SH2. (d) Semilog (base 10)
plots of probability
distributions $P(\delta v_{\alpha}(r))$ and $P(\delta b_{\alpha}(r))$, with $r$
in the
dissipation range; a Gaussian distribution is shown for comparison.}
\label{fig1}
\end{figure}
\vspace{-2.5cm}

\begin{figure}[htb]
\centerline{
\epsfxsize=9cm
\epsffile{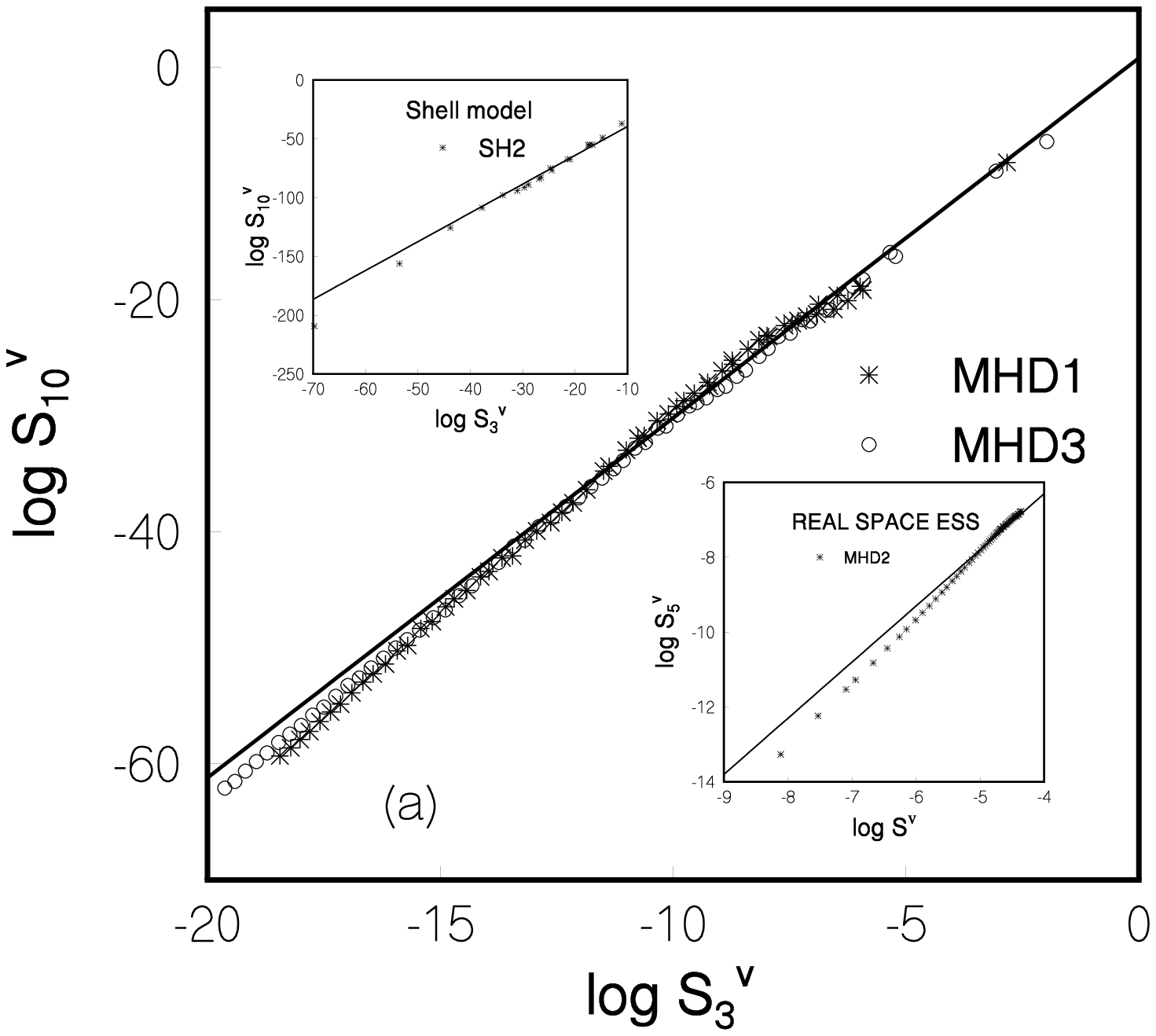}
\hfill
\epsfxsize=9cm
\epsffile{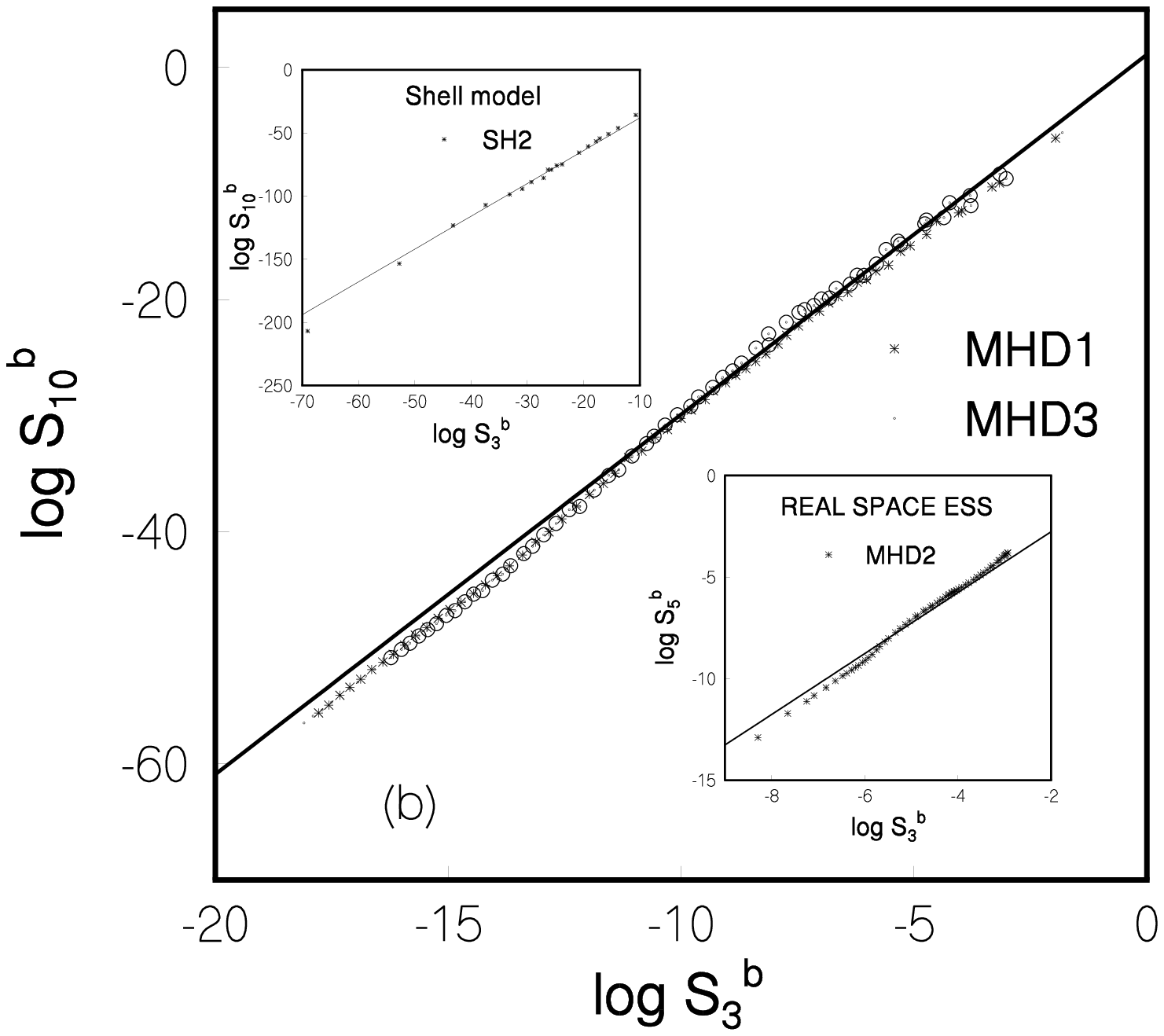}
}
\vspace*{-0.5cm}
\caption{Log-log plots (base 10) of $S_{10}^a$ versus $S_3^a$ showing $k$ space
ESS for $3d$MHD with (a) $a=v$ and (b) $a=b$. 
Insets illustrate real-space ESS for 
$3d$MHD and ESS for our shell model; the lines show the inertial-range 
asymptotes.} 
\label{fig3a}
\end{figure}
\begin{multicols}{2}
\noindent $\zeta^{'a}_p\/$ and $\alpha^a_p\/$
seem universal (the same for all our  
runs (Table 1));
$\alpha^a_p\/$ is close to, but {\em systematically less} than,
$p/3\/$.  
The $k\/$ dependences of $S_p ^a$ 
follow from that of $S^a_3\/$. 
We find 
\begin{eqnarray}
S^a_3 &\approx& B^a_I k^{-\zeta^a_3 - 9/2}, \;\; L^{-1} \ll k
\lesssim 1.5 k_d, \\
S^a_3 &\approx& B^a_D k^{\delta^a} \exp(-c^a k/k_d), \;\; 1.5
k_d \lesssim k \ll \Lambda,
\end{eqnarray}
where $B^a_I\/$ and $B^a_D\/$ are nonuniversal
amplitudes (Eq. (2) holds \cite{Dhar} for $3d$MHD; for our
shell model the factor $9/2\/$ is absent). Thus
{\em all} $S^a_p \sim k^{\theta^a_p} \exp(-
c^a \alpha^a_p k/k_d)\/$ for $1.5 k_d \lesssim k \ll \Lambda\/$,
with $\theta^a_p = \alpha^a_p \delta^a\/$
(cf. \cite{Dhar} for fluid turbulence). In Eq.(3)
$\delta^a, c^a\/$, and $k_d\/$ are not universal. However,
we extract the universal part of the inertial- to
dissipation-range crossover via our $k\/$-space GESS. 
We first define $G^a_{p} \equiv S^a_{p}/(S^a_{3})^{p/3}\/$;
log-log plots of $G^a_{p}\/$ versus $G^a_{q}\/$ yield curves 
with {\em universal, but different,}
slopes for asymptotes in inertial and dissipation ranges. The inertial-range
asymptote has a slope $\rho^a_{p,q}\/$ (as in real-space GESS
); 
the dissipation-range one has a slope $\omega^a(p,q)
\equiv [\alpha^a_p - p/3]/[\alpha^a_q - q/3]\/$. These slopes 
are universal, but not the points at which the
curves move away from the inertial-range asymptote. To obtain a {\em
universal crossover scaling function} (different for each
$(p,q)\/$ pair because of multiscaling) we
define
$\log(H^a_{pq}) \equiv D^a_{pq} \log(G^a_p)\/$ and
$\log(H^a_{qp}) \equiv D^a_{qp} \log(G^a_q)\/$; the scale
factors $D^a_{pq} = D^a_{qp}\/$ are {\em nonuniversal},
\vspace{-2.0cm}

\begin{figure}
\epsfysize=10cm
\centerline{\epsffile{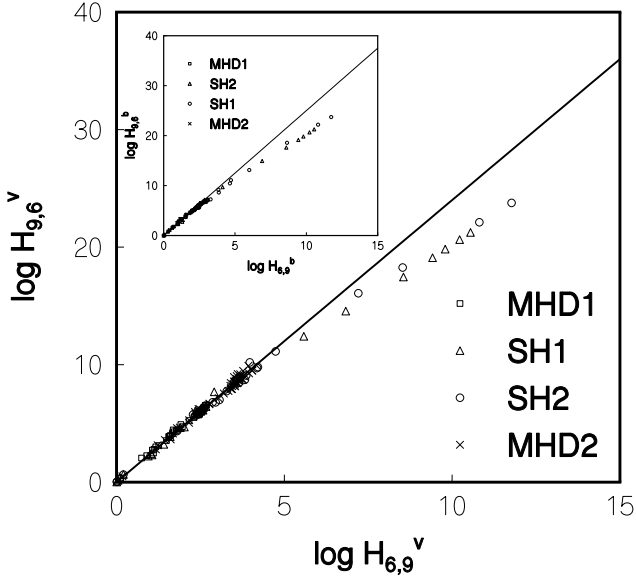}}
\caption{\narrowtext {Log-log plots (base 10) of 
$H_{9,6}^v$ versus 
$H_{6,9}^v$ and (inset)
$H_{9,6}^b$ versus $H_{6,9}^b$, illustrate our GESS
showing the universal inertial- to dissipation-range crossover; lines denote
inertial-range asymptotes.
}}
\label{fig3}
\end{figure}

\noindent but plots of
$\log(H^a_{pq})\/$ versus $\log(H^a_{qp})\/$ 
collapse onto a {\em  
universal curve} within our error bars 
for {\em all} $k\/$, $Re_{\lambda}\/$, and
$Re_{b\lambda}\/$ (Fig. 3).

The MHD equations
are \cite{mont}
\begin{equation}
{\partial {\bf{Z}}^{\pm} \over\partial t} + ({\bf Z^{\mp}} .{\bf 
\nabla)\bf Z^{\pm }} =
\nu_+ \nabla^2 {\bf Z^{\pm} } + \nu_- \nabla^2 {\bf Z^{\mp}} - {\nabla}p^* 
+ {\bf f^{\pm}},
\label{mhd1}
\end{equation}
where $\nu_{\pm} \equiv (\nu_v \pm \nu_b)/2\/$, $\nu_v\/$ and
$\nu_b\/$ are, respectively, the fluid and 
magnetic viscosities, $p^*$ 
is the effective pressure and the density $\rho=1$,
${\bf{f}}^{\pm} \equiv ({\bf{f}} \pm {\bf{g}})/2\/$, and 
${\bf{f}}\/$ and ${\bf{g}}\/$ are the forcing terms in the
equations for $\partial {\bf{v}}/\partial t\/$ and $\partial
{\bf{b}}/\partial t\/$. We assume incompressibility 
and use a pseudospectral method
\cite{Dhar} to solve 
Eq.(\ref{mhd1}) numerically. We force the first
two $k$-shells, use a cubical box with side $L_B = 2\pi$, periodic
boundary conditions, and $64^3\/$
modes in runs MHD1 and MHD2 and $80^3\/$ modes in run MHD3
(Table 1). We include fluid and magnetic
hyperviscosities (i.e., the term $-(\nu_v + \nu_{vH} 
k^2)k^2\/$ in the equation for $\partial {\bf{v}(\bf k)}/\partial t\/$ 
and the term $-(\nu_b + \nu_{bH} k^2)k^2$ in the equation for
$\partial {\bf{b}(\bf k)}/\partial t\/$, where $H\/$
stands for hyperviscosity). For time integration we
use an Adams-Bashforth scheme (step-size $\delta t$). We use
$Re_{\lambda}= v_{rms} \lambda /\nu_v$, $Re_{b\lambda}=
b_{rms}\lambda /\nu_b$, $\lambda_v =[\int_o ^\infty E_v
(k)dk/\int_o ^\infty k^2 E_v (k) dk]^{1/2}$, $\lambda_b =[\int_o
^\infty E_b (k)dk/\int_o ^\infty k^2 E_b (k) dk]^{1/2}$, $E_v
(k)\sim S_2 ^v (k) k^2$ and $E_b (k) \sim S_2 ^b(k) k^2$.
Parameters for runs MHD1-3 are given in Table
1, where $\tau_{ea} \equiv L_B/a_{rms}\/$ is the box-size
eddy-turnover time for field ${\boldmath{a}}\/$ and $\tau_{A}\/$ the
averaging time; initial transients are allowed to decay over a
period $\tau_t\/$. We use
quadruple-precision arithmetic; results
from our $64^3\/$ and $80^3\/$ runs are not significantly
different. 

Shell models for MHD turbulence have been proposed earlier
\cite{biskamp,carbone,gloa}, but there is {\em no}
MHD shell model that enforces {\em all} ideal $3d$MHD invariants
{\em and} which reduces to the GOY shell model for
fluid turbulence, when magnetic-field terms are supressed. We
present such a model and show that it yields
$\zeta_p ^a$
in agreement with those we obtain for
$3d$MHD. Our shell-model equations 
\begin{equation}
{dz_n ^{\pm }\over\ dt} = ic_n ^{\pm } -\nu_+ k_n^2 z_n
^{\pm } -\nu_-  k_n^2 z_n ^{\pm } + f_n ^{\pm } 
\label{mhd3}
\end{equation}
use the {\em complex, scalar}
Els\"{a}sser variables $z_n ^{\pm} \equiv (v_n \pm b_n)\/$, 
and discrete wavevectors $k_n=k_o q^n$, for
shells $n\/$; 

\begin{eqnarray}
c_n ^{\pm } &=& [a_1 k_n z_{n+1} ^{\mp } z_{n+2} ^{\pm }
+ a_2 k_n z_{n+1} ^{\pm } z_{n+2} ^{\mp } \nonumber \\
&+& a_3 k_{n-1} z_{n-1}^{\mp } z_{n+1} ^{\pm } 
+ a_4 k_{n-1} z_{n-1} ^{\pm } z_{n+1} ^{\mp } \nonumber \\ 
&+& a_5 k_{n-2} z_{n-1}^{\mp } z_{n-2}
^{\pm } + a_6 k_{n-2} z_{n-1} ^{\mp } z_{n-2} ^{\pm }]^* ,
\label{mhd2}
\end{eqnarray}
which ensures $z_n ^+ , z_n ^- \sim
k^{-1/3}\/$ is a stationary solution in
the inviscid, unforced limit \cite{biskamp}-\cite{kada} and preserves the
$\nu_+,Z^+ \leftrightarrow \nu_-,Z^-$ symmetry
of $3d$MHD. We fix five of the parameters, $a_1-a_6\/$,
by demanding that our shell-model analogs of the
total energy ($\equiv \sum_n (|v_n|^2 + |b_n|^2)/2\/$), the cross
helicity ($\equiv 1/2 \sum_n (v_n b_n^* + v_n^*b_n)$), and the magnetic helicity
($\equiv \sum_n (-1)^n |b_n|^2/k_n\/$) be conserved if
$\nu_{\pm} = 0\/$ and $f_n ^{\pm} = 0\/$; while enforcing the
conservation of energy, we also demand \cite{abasu} that the
cancellation of terms occurs as in
$3d$MHD.  We fix the last parameter
by demanding that, if $b_n = 0\/$ for all
$n\/$, our model should reduce to the GOY model, with the standard
choice of parameters \cite{kada} that conserves fluid helicity in the
inviscid, unforced limit. 
Thus, 
apart from the Reynolds numbers,
our shell model has no adjustable parameters and $a_1=7/12,
a_2=5/12, a_3=-1/12, a_4=-5/12, a_5=-7/12, a_6=1/12$, and $q =
2\/$. 
We solve Eq. (\ref{mhd3}) numerically by an
Adams-Bashforth scheme (step size $\delta t$), use 25
shells, force the first $k$-shell \cite{Dhar}, set
$k_o=2^{-4}$, and use $E_v =S_2 ^v (k_n)/k_n$, $\lambda _v=(2\pi/k_o)
[\Sigma_n S_{2}^v(k_n)/\Sigma_n k_n ^2 S_{2}^v(k_n)]^{1/2}$,
$\lambda_b=(2\pi/k_o) [\Sigma_n S_{2}^b(k_n)/\Sigma_n k_n ^2
S_{2}^b(k_n)]^{1/2}$, $v_{rms}=[k_o \Sigma_n S_{2}
^v(k_n)/\pi]^{1/2}$ and $b_{rms}=[k_o \Sigma_n S_{2}^b(k_n)
/\pi]^{1/2}$. Parameters for our four runs SH1-SH4
are given in Table 1. These use
double-precision arithmetic, but we have checked in
representative cases that our results are not affected if we use
quadruple-precision arithmetic.
As in the GOY model the structure functions 
$S_p(k_n)$ oscillate weakly with $k_n$
because of an underlying three-cycle \cite{kada,abasu}. These oscillations
can be removed either (a) by using ESS plots or (b) by using the structure
functions $\Sigma_{n,p}^a=\langle {\Im}[a_n a_{n+1} a_{n+2} + a_{n-1} a_n
a_{n+1}/4]^{p/3}\rangle$ \cite{kada}. Method (a) yields the exponent ratios
 $(\zeta_p ^a /\zeta_3 ^a)$, which we find are universal.
Method (b) gives exponents $\zeta^a_p\/$. These
have a mild dependence on $Re_{\lambda}\/$ and $Re_{b\lambda}\/$
but this goes away if we consider the exponent ratios 
$\zeta^a_p/\zeta^a_3\/$, as in the GOY model \cite{Dhar,lev}; thus
the asymptotes in our ESS and GESS plots
have universal slopes.

   The Navier Stokes equation ($3d$NS) follows from $3d$MHD if we set $\bf b=0$
or, equivalently, $Re_{b\lambda}$=0. However, if  
we start with $Re_{b\lambda} \simeq 0$, the steady state 
is characterised by the MHD exponents and
$Re_{\lambda}/Re_{b\lambda}\simeq
O(1)$ (i.e., an equipartition regime)\cite{foot4}. 
Since our MHD shell model reduces
to the GOY model as $Re_{b\lambda} \rightarrow 0$,
we use it to study the fluid turbulence to
MHD turbulence crossover, instead of doing costly 
pseudospectral studies: A small initial value of $Re_{b\lambda}$ yields
a transient during which we obtain GOY-model exponents, but
eventually the system crosses over to the MHD turbulence steady state
\cite{abasu}.

    In conclusion, then, we have shown 
that structure functions in $3d$MHD turbulence display multiscaling, ESS, and
GESS, with exponents and probability distributions $P(\delta 
v_{\alpha}({\bf r}))$
and $P(\delta b_{\alpha}({\bf r}))$  
different from those in 
fluid turbulence.  
Our new shell model (a) gives
the same multiscaling exponents 
as $3d$MHD and (b) reduces to the GOY shell model as 
$Re_{b\lambda} \rightarrow 0$. Our ESS and GESS studies help
us to uncover an apparently universal crossover from inertial- to
dissipation-range asymptotics. It would be very interesting to compare
our results with experiments on MHD turbulence, but two points must be borne
in mind: (1) solar-wind data might yield multiscaling
exponents different from ours because of the presence of a mean magnetic field;
(2) the crossover from inertial- to dissipation-range asymptotics 
might not apply to the solar wind because a hydrodynamic
description might break down in the dissipation range \cite{marsch}. 
However, our results should apply to  MHD systems which show an equipartition
regime \cite{mont}.
It would also be
interesting to see whether the agreement of
 $\zeta_p ^b$ with the SL formula is fortuitous or significant.

We thank J.K. Bhattacharjee and S. Ramaswamy for discussions, 
CSIR (India) for
support, and SERC (IISc, Bangalore) for computational resources.

\end{multicols}
\newpage

\begin{table}[t]
\widetext
\caption{The viscosities and hyperviscosities $\nu_v, \nu_b,
\nu_{vH}\/$ and $\nu_{bH}\/$, the Taylor-microscale Reynolds numbers
$Re_{\lambda}\/$ and $Re_{b\lambda}\/$, the box-size eddy-turnover
times $\tau_{ev}\/$ and $\tau_{eb}\/$, the averaging time
$\tau_{A}\/$, the time over which transients are allowed to
decay $\tau_t\/$, and $k_d\/$ (dissipation-scale wavenumber) for
our $3d$MHD runs ($k_{max} = 32\/$ for MHD1 and MHD2 and
$k_{max} = 40\/$ for MHD3) and shell-model runs SH1-4 ($k_{max}
= 2^{25} k_0\/$). The step size($\delta t\/$) is 0.02 for
MHD1-3, $2.10^{-5}$ for SH1-2, and 
$10^{-4}\/$ for SH3-4. Note that $\tau_{ev} \simeq
8\tau_I\/$ the integral time for our MHD runs.} 
\label{table1}
\begin{tabular}{|l|c|c|c|c|c|c|c|c|c|c|c|}
Run & $\nu_v$ & $\nu_{vH}$ & $\nu_b$ & $\nu_{bH}$ &
$Re_{\lambda}$ & $Re_{m\lambda}$ & $\tau_{ev}/\delta t\/$ &
$\tau_{ev}/\delta t\/$ & $\tau_t/\tau_{ev}\/$ & $\tau_{A}/\tau_{ev}\/$ &
$k_{max}/k_d\/$ \\ \hline 
MHD1 & $8 \cdot 10^{-4}$ & $7 \cdot 10^{-6}$ &$ 10^{-3}$
& $8 \cdot 10^{-6}$ &  $\simeq 24.8$ &   $\simeq 14.3$ & 
$\simeq 8.8 \cdot 10^3\/$ &$\simeq 6 \cdot 10^3\/$ & $\simeq 2
\/$ &$\simeq 2.3 \/$ &  $\simeq 1.83\/$ \\
MHD2 & $8 \cdot 10^{-4}$ & $9 \cdot 10^{-6}$ & $8 \cdot 10^{-4}$
& $9 \cdot 10^{-6}$ &  $\simeq 24.1$ &   $\simeq 18.1$ & 
$\simeq 8.8 \cdot 10^3\/$ & $\simeq 5.6 \cdot 10^3\/$ & $\simeq 2
\/$ & $\simeq 2.3 \/$ &  $\simeq 1.83\/$ \\
MHD3 & $8 \cdot 10^{-4}$ & $9 \cdot 10^{-6}$ & $8 \cdot 10^{-4}$
& $9 \cdot 10^{-6}$ &  $\simeq 26$ &   $\simeq 19.6$ & 
$\simeq 7.9 \cdot 10^3\/$ & $\simeq 4.8 \cdot 10^3\/$ & $\simeq 1
\/$ & $\simeq 2.2 \/$ &  $\simeq 2.22\/$ \\
SH1 & $10^{-9}$ & $0$ & $10^{-9}$
& $0$ &  $\simeq 4.6 \cdot 10^8$ &   $\simeq 7.8 \cdot 10^8$ & 
$\simeq 10^7\/$ & $\simeq 6\cdot 10^6\/$ & $\simeq 50
\/$ & $\simeq 450\/$ &  $\simeq 2^5\/$ \\
SH2 & $10^{-8}$ & $0$ & $10^{-8}$
& $0$ &  $\simeq 4.3 \cdot 10^7$ & $\simeq 6.5 \cdot 10^7$ & 
$\simeq 10^7\/$ & $\simeq 6\cdot 10^6\/$ & $\simeq 50
\/$ & $\simeq 450\/$ &  $\simeq 2^8\/$ \\
SH3 & $10^{-6}$ & $0$ & $2 \cdot 10^{-6}$
& $0$ &  $\simeq 4 \cdot 10^6$ & $\simeq 3 \cdot 10^6$ & 
$\simeq 2 \cdot 10^6\/$ & $\simeq 10^6\/$ & $\simeq 500
\/$ & $\simeq 2500\/$ &  $\simeq 2^{10}\/$ \\
SH4 & $4 \cdot 10^{-6}$ & $0$ & $10^{-6}$
& $0$ &  $\simeq 1.2 \cdot 10^5$ & $\simeq 1 \cdot 10^6$ & 
$\simeq 10^6\/$ & $\simeq 1.7 \cdot 10^6\/$ & $\simeq 500
\/$ & $\simeq 3000\/$ &  $\simeq 2^{11}\/$ \\
\end{tabular}
\end{table}

\begin{references}
\bibitem{foot3} We use the SL formula as a convenient parametrization for
the multiscaling exponents in fluid turbulence.
\bibitem{foot4} Thus in a 
renormalization-group calculation $Re_{b\lambda}$ should appear as
a relevant operator that takes the system from the $3d$NS-turbulence fixed
point to the $3d$MHD-turbulence fixed point.
\end{references}
\end{document}